\documentclass[twocolumn]{aastex701}

\usepackage{amsmath,amssymb,amsfonts}
\usepackage{booktabs}
\usepackage{microtype}
\usepackage{graphicx}
\usepackage{url}
\usepackage{xcolor}

\newcommand{\tic}{TIC~198456033}
\newcommand{\hdstar}{HD~158485}
\newcommand{\porb}{P_{\rm orb}}

\shorttitle{A 383-day PB2 Orbit in HD 158485}
\shortauthors{Lv et al.}

\begin{document}

\title{TESS Pulsation Timing Reveals the JWST Calibration Star HD 158485 as a Double-pulsator Binary}

%\title{The JWST Calibration Star HD 158485 Is a 386-day Double-pulsator Binary}

\author[orcid=0000-0001-6354-1646]{Chenglong Lv}
\affiliation{Xinjiang Astronomical Observatory, Chinese Academy of Sciences, Urumqi, Xinjiang 830011, People's Republic of China}
\affiliation{Instituto de Astrof\'isica de Andaluc\'ia - CSIC, 18008 Granada, Spain}
\email[show]{lvchenglong@xao.ac.cn}

\author[orcid=0000-0003-1845-4900]{Ali Esamdin}
\affiliation{Xinjiang Astronomical Observatory, Chinese Academy of Sciences, Urumqi, Xinjiang 830011, People's Republic of China}
\affiliation{School of Astronomy and Space Science, University of Chinese Academy of Sciences, Beijing 100049, People's Republic of China}
\email{aliyi@xao.ac.cn}

\author[orcid=0000-0001-5066-5682]{Shuguo Ma}
\affiliation{Xinjiang Astronomical Observatory, Chinese Academy of Sciences, Urumqi, Xinjiang 830011, People's Republic of China}
\email{mashuguo@xao.ac.cn}

\author[0000-0001-8677-2391]{Shahidin Yaqup}
\affiliation{Xinjiang Astronomical Observatory, Chinese Academy of Sciences, Urumqi, Xinjiang 830011, People's Republic of China}
\affiliation{School of Astronomy and Space Science, University of Chinese Academy of Sciences, Beijing 100049, People's Republic of China}
\email{xiayiding@xao.ac.cn}

\author[orcid=0000-0002-2463-4943]{Haozhi Wang}
\affiliation{Chinese Academy of Sciences South America Center for Astronomy}
\email{wanghaozhi@xao.ac.cn}

\author[orcid=0000-0001-9600-963X]{Ayzada Jmahali}
\affiliation{Xinjiang Astronomical Observatory, Chinese Academy of Sciences, Urumqi, Xinjiang 830011, People's Republic of China}
\affiliation{School of Astronomy and Space Science, University of Chinese Academy of Sciences, Beijing 100049, People's Republic of China}
\email{Ayzada@xao.ac.cn}

\author[orcid=0009-0000-2772-2677]{Jianwei Yao}
\affiliation{Xinjiang Astronomical Observatory, Chinese Academy of Sciences, Urumqi, Xinjiang 830011, People's Republic of China}
\affiliation{School of Astronomy and Space Science, University of Chinese Academy of Sciences, Beijing 100049, People's Republic of China}
\email{yaojianwei@xao.ac.cn}

\author[orcid=0009-0008-7094-1304]{Kedieryemu Kamili}
\affiliation{School of Physical Science and Technology, Xinjiang University, Urumqi 830017, People's Republic of China}
\email{kedieryemu@xao.ac.cn}

\begin{abstract}
HD~158485 is an A-type $\delta$~Scuti star in the CALSPEC network and the JWST absolute-flux calibration program. Gaia DR3 assigns it a binary classification probability close to unity
but provides no orbital solution. Using TESS observations from Sectors 14--26, 40--41, 47--60, and 73--86, we measure the pulsation phases of six $\delta$~Scuti modes. The frequencies form two groups with opposite light-travel-time delays, revealing two pulsating components. A direct light-curve forward model yields an orbital period of $\porb=382.80^{+0.80}_{-0.76}$~d, eccentricity $e=0.242^{+0.045}_{-0.046}$, projected light-time amplitudes of $195.8^{+9.2}_{-8.9}$ and $239.7^{+6.8}_{-6.8}$~s, and a dynamical mass ratio of $q=M_B/M_A=0.818^{+0.046}_{-0.045}$. A blind prediction based only on the first two long TESS campaigns reproduces the third campaign without updating the orbital parameters, with a Pearson correlation coefficient of $r=0.983$. Its held-out rms is 83.8~s, compared with 289.9~s when the period is fixed at 365.25~d. A 5.15-$\sigma$ Hipparcos--Gaia EDR3 proper-motion anomaly independently indicates unresolved astrometric motion. The unresolved companion can also affect spectrophotometric predictions: a conditional PB2-constrained two-atmosphere model differs from the corresponding single-star atmosphere prediction by 1.6\% over 1--5~$\mu$m and 2.5\% over 5--28~$\mu$m. HD~158485 therefore shows that a calibration-relevant binary can lack a catalogued orbital solution while two independent pulsation clocks reveal and measure its orbit.
\end{abstract}


\section{Introduction}\label{sec:intro}

Absolute flux calibration relies on accurate stellar spectral predictions. The James Webb Space Telescope (JWST) calibration program therefore uses multiple calibrators, including hot stars, A dwarfs, and solar analogs, to establish an absolute flux scale and evaluate uncertainties in stellar atmosphere models \citep{Gordon2022}. The selection criteria favor single stars or wide binaries because unresolved companions can introduce composite spectra requiring more complex modeling. Unresolved multiplicity can nevertheless remain difficult to characterize when no orbital solution is available.

\hdstar\ (HR~6514, \tic) is an A4~V star included in the CALSPEC database and the JWST A-dwarf calibration network \citep{Bohlin2017,Gordon2022}. Its short-period variability was identified as $\delta$~Scuti pulsation from TESS observations \citep{Mullally2022b}. That study did not report an orbital or PB2 interpretation.

Pulsating stars can also act as precise orbital clocks. Orbital motion changes pulsation phases through the light-travel-time effect \citep{ShibahashiKurtz2012,Murphy2014}. Frequencies from the same star share a common delay curve, whereas double-pulsator binaries (PB2s) exhibit two coherent timing signals with opposite light-travel-time variations produced by a shared orbit \citep{Murphy2018}. Although pulsation timing has revealed many binaries \citep{Murphy2018,2026arXiv261000062L}, PB2 systems remain rare \citep{Dholakia2025}. 

Previous observations had already provided evidence for unresolved multiplicity, although no orbit was available. Speckle observations did not resolve the system \citep{McAlister1987}. In Gaia DR3, the Discrete Source Classifier gives a combined-model binary-star probability of 0.99999744 \citep{Delchambre2023}. The Multiple Star Classifier gives conditional component temperatures of $7600^{+53}_{-19}$ and $7599^{+36}_{-50}$~K, with both surface
gravities close to $\log g=3.90$, under its unresolved-binary model
\citep{Creevey2023}. The source has a RUWE of 1.1366 but no published Gaia DR3 non-single-star orbital solution \citep{GaiaDR3}. A 5.15-$\sigma$ Hipparcos--Gaia EDR3 proper-motion anomaly also indicates unresolved astrometric motion \citep{Kervella2022}. Table~\ref{tab:target} summarizes the basic properties of HD~158485.

Here we use the extended TESS baseline to reveal that the pulsations of \hdstar\ contain two independent orbital clocks. Six pulsation frequencies reveal two timing groups with opposite
light-travel-time variations that trace a common orbit of approximately 383 days, allowing the projected orbital motions of both components and the dynamical mass ratio to be constrained. We validate the orbit through a blind inter-campaign prediction test using only earlier observations. We further examine whether a physically consistent two-star model can remain consistent with the optical CALSPEC spectrum while producing measurable differences in infrared flux predictions.

\begin{deluxetable}{ll}
\tabletypesize{\small}
\tablecaption{Basic Properties of HD 158485\label{tab:target}}
\tablewidth{0pt}
\tablehead{
\colhead{Quantity} &
\colhead{Value}
}
\startdata
Gaia DR3 source ID & 1434752727380965760 \\
Spectral type & A4~V \\
$G$ & 6.4763 mag \\
$G_{\rm BP}-G_{\rm RP}$ & 0.19045 mag \\
Parallax & 8.3579 mas \\
RUWE & 1.1366 \\
DSC binary probability & 0.99999744 \\
MSC $T_{{\rm eff},1}$ & $7600^{+53}_{-19}$ K \\
MSC $T_{{\rm eff},2}$ & $7599^{+36}_{-50}$ K \\
Integrated $T_{\rm eff}$ & 8044 K \\
Single-star-equivalent mass & $2.38\pm0.12\,M_\odot$ \\
PMA significance  & $5.15\,\sigma$ \\
\enddata
\tablecomments{Gaia astrometry and photometry are from Gaia DR3 \citep{GaiaDR3}; the DSC and MSC quantities are from \citet{Delchambre2023} and \citet{Creevey2023}, respectively. The parallax is from the five-parameter astrometric solution. The DSC value is a classification probability rather than an orbital solution, and the MSC component parameters are conditional on the Gaia unresolved-binary model. The spectral type and integrated $T_{\rm eff}$ are from \citet{Su2006}; the single-star-equivalent mass and proper-motion anomaly (PMA) significance are from \citet{Kervella2022}.}
\end{deluxetable}

\section{TESS Photometry and Pulsation Timing}\label{sec:data}

We used two-minute cadence TESS photometry from Sectors 14--26, 40--41, 47--60, and 73--86 \citep{Ricker2015}. The observations consist of three extended campaigns (Sectors 14--26, 47--60, and 73--86) and one additional short visit (Sectors 40--41), providing a total baseline of approximately 2000~d. We adopted the SPOC PDCSAP light curves, removed flagged cadences and non-finite measurements, and normalized each sector independently before combining the data.

We calculated the Lomb--Scargle amplitude spectrum of the combined TESS light curve and identified six isolated $\delta$~Scuti modes near 15--19~d$^{-1}$ \citep{Lomb1976,Scargle1982}. The selected frequencies were chosen based on their amplitudes, spectral isolation, and suitability for phase measurements over the full TESS baseline. Three frequencies ($17.792354$, $17.945470$, and $15.127640$~d$^{-1}$) form one timing group, while three frequencies ($18.619852$, $18.042427$, and $19.254772$~d$^{-1}$) form the second group. 

\begin{figure}
 \centering
 \includegraphics[width=0.45\textwidth]{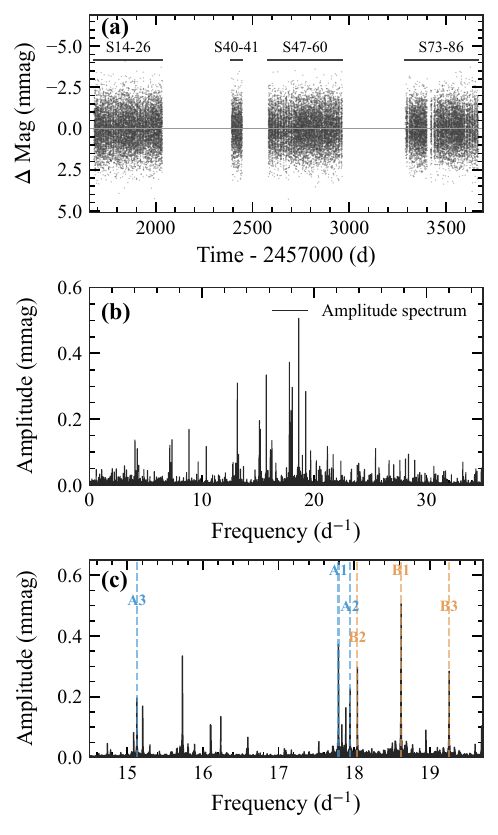}
\caption{TESS photometry and pulsation-mode selection.
(a) Differential magnitude in mmag and observing windows. Sectors 14--26, 47--60, and 73--86 correspond to the three extended campaigns, while Sectors 40--41 provide an additional short visit.
(b) Amplitude spectrum of the full TESS baseline.
(c) The six pulsation frequencies used for timing analysis. The two colors indicate the two timing groups adopted in this work.}
 \label{fig:overview}
\end{figure}

For repeated likelihood calculations, adjacent samples within each continuous segment were averaged to a 20-min cadence, leaving 68,153 flux measurements. These retained flux measurements were used in the direct light-curve forward model. The 10-min series was used for the segmented phase measurements employed for visualization and validation; these measurements were not included as data in the forward-model likelihood. This resampling has a negligible effect on the timing result: the delay curves from the 20-min and 10-min analyses agree with Pearson correlations of 0.99884--0.99992 and median absolute differences of 1.7--5.9~s. \citet{Mullally2022b} found no evidence that the variability of \hdstar\ was caused by nearby contamination and reported a CROWDSAP value of 1.000 for the TESS photometry. The lack of evidence for contamination supports the association of the detected pulsations with the target system. Figure~\ref{fig:overview} shows the TESS observing windows, amplitude spectrum, and the six pulsation modes used in the timing analysis. The two colors indicate the two timing groups identified from the measured light-travel-time variations.

The primary orbital inference was obtained by forward modeling the light-travel-time variations directly in the 68,153 retained flux measurements. For mode $k$ belonging to component $j$, the model is

\begin{equation}
\begin{aligned}
y(t) ={}& \mu+
\sum_{j\in\{A,B\}}\sum_{k\in j}
\bigl[c_k\cos\Theta_{jk}(t)\\
&\qquad\qquad
+s_k\sin\Theta_{jk}(t)\bigr],\\
\Theta_{jk}(t) ={}&
2\pi\nu_k\left[t-\tau_j(t)\right].
\end{aligned}
\label{eq:forward}
\end{equation}

Here $\tau_A(t)$ and $\tau_B(t)$ are Keplerian light-travel-time curves that share the orbital period, eccentricity, argument of periastron, and
orbital phase, but have opposite signs and separate projected light-time amplitudes. The twelve sine and cosine coefficients were solved in a single
linear least-squares step at each orbital evaluation. The likelihood was evaluated directly on the flux series. We modeled the light curve following the forward-modeling method introduced by \citet{Hey2020Forward}, using its implementation in \texttt{Maelstrom} \citep{Hey2020}.

The centers of the priors on $P$, $a_A\sin i/c$, and $a_B\sin i/c$ were taken from a preliminary fit to the segmented delays, while the prior widths were chosen to be substantially broader than the corresponding segmented-fit uncertainties. We adopted $\log(P/{\rm d})\sim\mathcal{N}[\log(386.47),0.08]$,
truncated to $300<P/{\rm d}<480$, uniform priors on orbital phase and $\omega$ over $[0,2\pi)$, and $e\sim{\rm Beta}(1.5,4)$. The light-time amplitudes had $\log(a_A/{\rm s})\sim\mathcal{N}[\log(195.59),0.35]$ and $\log(a_B/{\rm s})\sim\mathcal{N}[\log(226.95),0.35]$. Each pulsation frequency had a normal prior of width $10^{-5}\,{\rm d}^{-1}$ about its full-baseline value. The relative-flux mean had a $\mathcal{N}(0,1)$ prior, and the logarithmic residual scatter had a normal prior of width 2 about the logarithm of the sample standard deviation. The twelve linear sine and cosine coefficients were profiled rather than sampled.

Four independent chains with 10,000 retained draws each gave $\hat R\leq1.01$ and bulk effective sample sizes $\geq483$ for the orbital parameters. The approximately 40-d segmented delays were used for visualization, comparison with the forward-model result, and the independent inter-campaign prediction test in Section~\ref{sec:validation}.

\section{The PB2 Orbit}\label{sec:orbit}

\begin{figure*}
 \centering
 \includegraphics[width=0.75\textwidth]{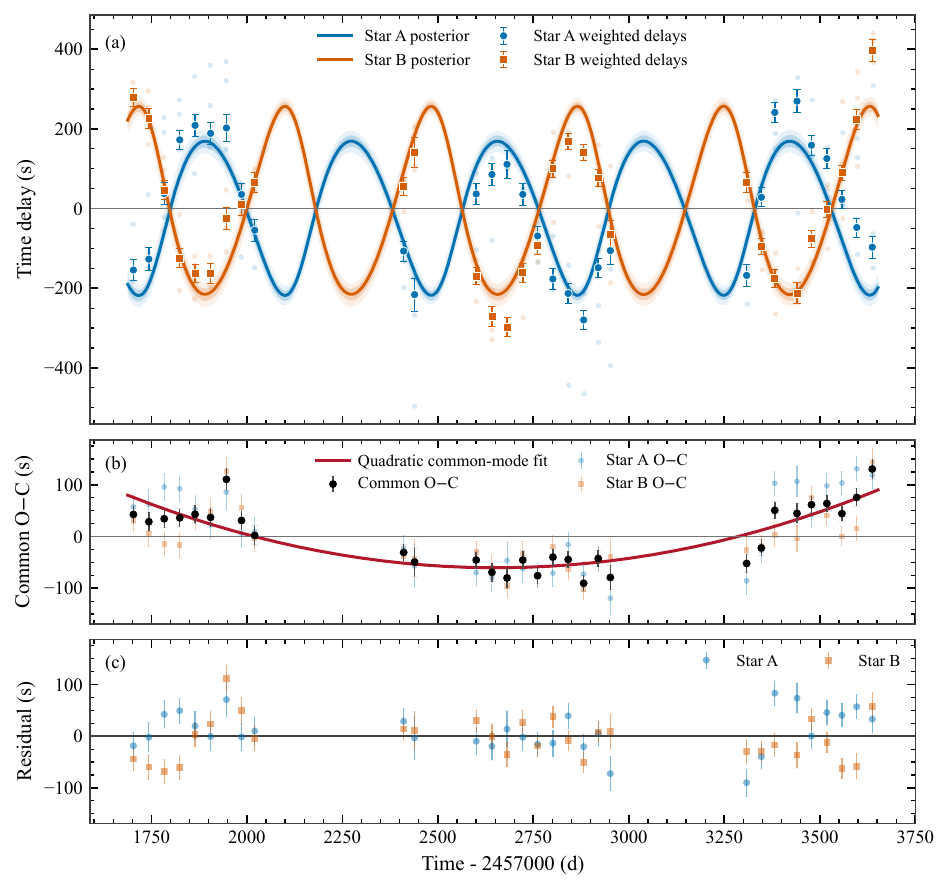}
 \caption{PB2 pulsation-timing solution.
(a) Individual 40-d delays (pale symbols), residual-scatter-weighted component delays (dark symbols), and predictions from the forward-model posterior. Error bars are propagated from the individual phase fits. The segmented points are not fitted in deriving the curves.
(b) The common O-C residual and a phenomenological quadratic description.
(c) Component residuals after subtracting the PB2 orbit and common quadratic. The common trend is not interpreted as a third body.}
 \label{fig:delays}
\end{figure*}

For presentation, the three segmented mode delays in each timing group were combined using residual-scatter weighting to construct the component delay series $\tau_A$ and $\tau_B$ shown in Figure~\ref{fig:delays}. The two groups trace opposite light-travel-time variations produced by a common orbit. The segmented delays are shown for comparison with the forward-model predictions.

The forward model gives an orbital period of $382.80^{+0.80}_{-0.76}$~d and an eccentricity of $0.242^{+0.045}_{-0.046}$. The projected light-time amplitudes are $195.8^{+9.2}_{-8.9}$~s and $239.7^{+6.8}_{-6.8}$~s, giving a dynamical mass ratio of $q=M_B/M_A=a_A/a_B=0.818^{+0.046}_{-0.045}$. A separate fit to the segmented delays gave $386.5\pm1.9$~d and compatible amplitudes and eccentricity. The two analyses use different representations of the same photometry: the segmented fit compresses each 40-d interval into phase measurements, whereas the forward model evaluates the likelihood directly from all retained flux measurements. We therefore use the segmented result as a diagnostic and adopt the forward-model result as
the orbital solution.

The forward model leaves a residual curvature shared by both timing groups (Figure~\ref{fig:delays}). This common O-C signal is distinct from the anti-phase PB2 orbital signal and is therefore considered separately from the orbital solution. If interpreted entirely as secular pulsation evolution, it would imply $\dot P/P\simeq1.3\times10^{-6}$~yr$^{-1}$, larger than typical values expected for slowly evolving main-sequence pulsators \citep{Breger1998}. Moreover, such an interpretation would require independent pulsation
modes from the two timing groups to share a common period drift. The common curvature was not included in the adopted full-light-curve likelihood. We treat it only as a descriptive residual trend and do not interpret it as evidence for an additional companion.

Table~\ref{tab:orbit} summarizes the fitted and derived orbital quantities. The light-time amplitudes correspond to projected barycentric semimajor axes of 0.392 and 0.480~au. The predicted radial-velocity semiamplitudes are $11.51^{+0.56}_{-0.56}$ and $14.07^{+0.50}_{-0.47}$~km~s$^{-1}$, suggesting that future spectroscopic verification is feasible, although the timing identification does not rely on spectroscopy.

\begin{deluxetable*}{lccc}
\tablecaption{PB2 Orbital and Derived Parameters\label{tab:orbit}}
\tablehead{\colhead{Parameter} & \colhead{Median} & \colhead{$-1\sigma$} & \colhead{$+1\sigma$}}
\startdata
$\porb$ (d) & 382.80 & 0.76 & 0.80\\
$a_A\sin i/c$ (s) & 195.8 & 8.9 & 9.2\\
$a_B\sin i/c$ (s) & 239.7 & 6.8 & 6.8\\
$e$ & 0.242 & 0.046 & 0.045\\
$\omega$ (deg) & 129.0 & 10.2 & 10.1\\
$q=M_B/M_A$ & 0.818 & 0.045 & 0.046\\
$K_A$ (km s$^{-1}$) & 11.51 & 0.56 & 0.56\\
$K_B$ (km s$^{-1}$) & 14.07 & 0.47 & 0.50\\
$f_A(M)$ ($M_\odot$) & 0.0551 & 0.0073 & 0.0080\\
$f_B(M)$ ($M_\odot$) & 0.1009 & 0.0084 & 0.0089\\
$(M_A+M_B)\sin^3i$ ($M_\odot$) & 0.605 & 0.046 & 0.049
\enddata
\tablecomments{Quoted intervals correspond to the posterior 16th--84th percentiles. Component labels refer to the two timing groups and not to brightness ordering.}
\end{deluxetable*}

\section{A Blind Inter-campaign Prediction}\label{sec:validation}

The orbital period is close to the duration of a 14-sector TESS campaign. We therefore performed a predictive test to distinguish a coherent orbital signal from a possible campaign-dependent systematic effect. The differential delay, $D=\tau_A-\tau_B$, was fitted using only Sectors 14--26 and 47--60. Sectors 73--86 were excluded from the fit, and the resulting model was
propagated forward without updating any fitted parameter. For unequal orbital amplitudes, the differential delay can be written as
\begin{equation}
\begin{aligned}
\tau_A &= A_A g(t)+C(t)+n_A(t),\\
\tau_B &=-A_B g(t)+C(t)+n_B(t),\\
D&=(A_A+A_B)g(t)+n_A(t)-n_B(t).
\end{aligned}
\label{eq:differential}
\end{equation}
Here $A_A$ and $A_B$ are the positive orbital light-time amplitudes, $g(t)$ is the common orbital shape, $C(t)$ is any delay shared by both timing groups, and $n_A$ and $n_B$ collect group-specific noise. Thus $C(t)$ cancels exactly; unequal $A_A$ and $A_B$ add in the differential orbital signal rather than leaving a common-mode residual.

\begin{figure}
 \centering
 \includegraphics[width=0.45\textwidth]{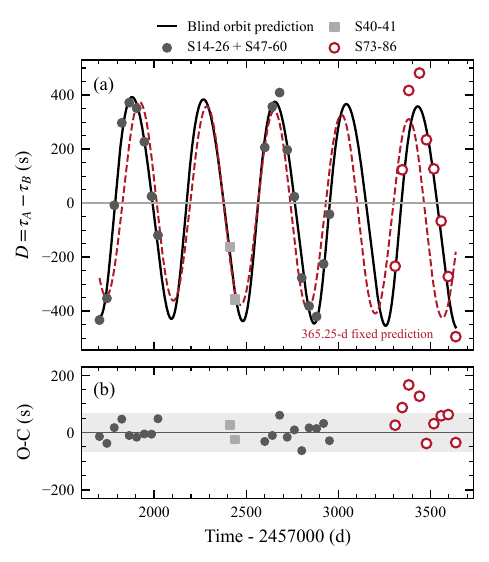}
 \caption{Blind differential-delay prediction. Filled gray points show the training data used to fit the orbit, gray squares show S40--41 excluded from the fit, and open red circles show the independent S73--86 test data. The black curve shows the free-period prediction without refitting any orbital parameter. The red dashed curve is the otherwise identical prediction with $P$ fixed at 365.25~d. The bottom panel shows residuals for all campaigns; the gray band indicates the global $1\sigma$ differential-delay scale.}
 \label{fig:blind}
\end{figure}

The training-only solution gives $\porb=387.70$~d and $e=0.201$. When applied to the independent Sectors 73--86 data, the predicted and measured differential delays show a Pearson correlation of $r=0.983$ and a root-mean-square error of 83.8~s. We repeated the same training-only fit with $P$ fixed at exactly 365.25~d. The free-period and fixed-period models used the same parameterization, except that $P$ was fixed in the latter.
In both cases, $e$, $\omega$, phase, differential amplitude, offset, and linear drift were free. No parameter was updated on Sectors 73--86. The fixed-year model has a held-out rms of 289.9~s, 3.46 times larger than the free-period result, and its held-out correlation falls to 0.555. Figure~\ref{fig:blind} shows the held-out measurements and the prediction obtained without refitting any orbital parameter.

A campaign-dependent systematic effect would need to reproduce not only an approximately annual timescale, but also the phase of the differential-delay curve after multi-year extrapolation and the opposite timing response of the two timing groups. The blind prediction therefore makes a purely instrumental origin unlikely. 

Additional checks using different segment lengths, SAP and PDCSAP products, nearby control stars, and campaign-by-campaign anti-phase measurements are presented in Appendix~\ref{app:validation}.

\section{Physical and Calibration Implications}\label{sec:physical}

The timing orbit provides physical constraints even without radial velocities. Combining the two projected semimajor axes gives the projected total mass:
\begin{equation}
 (M_A+M_B)\sin^3i=0.605^{+0.049}_{-0.046}\,M_\odot .
\end{equation}

The $2.38\pm0.12\,M_\odot$ value reported by \citet{Kervella2022} was derived under a single-star assumption and therefore cannot be assigned directly to either timing component. We use it only as an unresolved, single-star-equivalent mass scale. Combining this scale with the empirical main-sequence relations of \citet{Eker2018} and the measured mass ratio provides a consistency estimate of $M_A\simeq2.0$--$2.2\,M_\odot$ and $M_B\simeq1.7$--$1.9\,M_\odot$. These estimates imply a possible inclination of $\sim30^\circ$ and a relative semimajor axis of $\sim1.6$~au, but remain conditional on the adopted external mass scale. We therefore quote only timing-derived quantities in Table~\ref{tab:orbit}.

The relatively high mass ratio also helps explain why the companion is not conspicuous in the integrated spectrum. In the relevant mass interval, the empirical relation is approximately $L\propto M^{4.329}$ \citep{Eker2018}, giving $L_B/L_A\simeq q^{4.329}\simeq0.42$ for the posterior median mass ratio (approximately 0.33--0.53 over the quoted $q$ interval). The corresponding empirical temperature scaling gives $T_B/T_A\simeq0.85$. We used these relations to construct a conditional two-atmosphere prediction with the updated BOSZ grid \citep{Meszaros2024}. The two components were assigned the same composition, extinction, and distance, while the timing-derived mass-ratio posterior was propagated into their temperatures, radii, and gravities. With the forward-model mass ratio, a representative realization has $M_A=2.20\,M_\odot$, $M_B=1.80\,M_\odot$, $T_A=9010$~K, and $T_B=7690$~K. These parameters provide a conditional decomposition for testing the wavelength dependence of the unresolved flux. This conditional decomposition is not based on the Gaia MSC solution, which uses different data and model assumptions.

At the Gaia DR3 parallax of 8.3579~mas \citep{GaiaDR3}, the conditional semimajor axis corresponds to approximately 13--14~mas, and the expected apastron separation is about 16--17~mas. Both are well below the 38--40~mas resolution limit of the \citet{McAlister1987} speckle survey, making the non-detection expected. The target's RUWE of 1.1366 is below the RUWE \(>1.4\) threshold used in the initial selection of sources for the Gaia DR3 astrometric-binary processing \citep{Halbwachs2023}. In addition, the near-annual orbital period is susceptible to covariance with parallax, while the comparable fluxes of the two components further suppress the photocentre motion. These effects may help explain the absence of a published Gaia DR3 NSS orbit.

The 5.15-$\sigma$ Hipparcos--Gaia EDR3 proper-motion anomaly indicates unresolved astrometric motion but does not by itself identify the responsible orbit. A photocentre consistency calculation using the inner PB2 orbit, the conditional total-mass range of $3.7$--$4.1\,M_\odot$, and the optical light ratio from the two-atmosphere model gives $0.357^{+0.090}_{-0.098}$~mas~yr$^{-1}$, comparable to $0.284\pm0.055$~mas~yr$^{-1}$ corresponding to the $161\pm31$~m~s$^{-1}$ proper-motion anomaly reported by \citet{Kervella2022}. The inner PB2 can therefore plausibly account for the observed anomaly, and the proper-motion anomaly alone does not require an additional wider companion.

\begin{figure}
 \centering
 \includegraphics[width=0.45\textwidth]{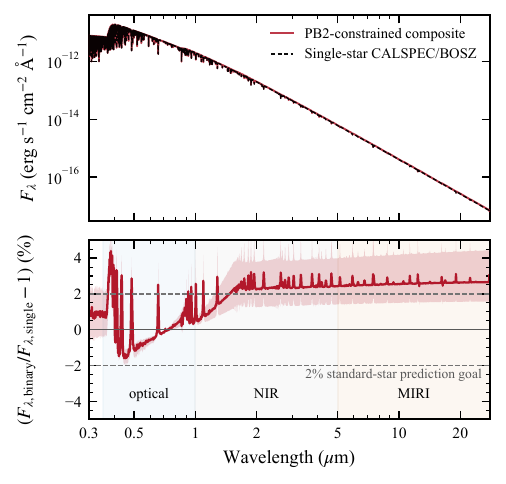}
 \caption{Conditional composite-spectrum test. The upper panel compares the CALSPEC spectrum with PB2-constrained composite and single-star BOSZ models. The lower panel shows the binary-to-single flux ratio and its uncertainty from the timing mass ratio. The optical region was used for normalization; no additional NIR or MIRI scaling was applied. Dashed lines indicate $\pm2\%$.}
 \label{fig:calspec}
\end{figure}

Using the same two-atmosphere model, we compared the composite spectrum with a single-star BOSZ model corresponding to the CALSPEC atmospheric parameters and applied the differential ratio to the delivered CALSPEC
flux distribution \citep{Bohlin2014,Bohlin2017}. Figure~\ref{fig:calspec} shows the comparison. Only one normalization was fitted over 0.35--1.0~$\mu$m; no additional scaling was applied in the near-infrared (NIR) or Mid-Infrared Instrument (MIRI) ranges. The band-integrated binary-minus-single difference in the optical is $-0.004\%$, with a 16th--84th percentile interval of $-0.032\%$ to $+0.015\%$. The rms optical spectral-shape difference is 1.26\%, because the normalization removes only the mean flux offset and not wavelength-dependent structure. The corresponding offsets are $+1.62\%$ ($+0.92\%$ to $+2.65\%$) over 1--5~$\mu$m and $+2.53\%$ ($+1.48\%$ to $+4.14\%$) over 5--28~$\mu$m.

The median NIR shift is smaller than the nominal 2\% prediction accuracy sought for an individual JWST standard, whereas the median MIRI shift is slightly larger \citep{Gordon2022}. This comparison is conditional on the empirical mass--temperature--radius relations and is not evidence for a JWST zeropoint error: the calibration program combines multiple standards, and the calculation tests a binary-versus-single spectral prediction rather than the delivered calibration itself. It demonstrates quantitatively why an unresolved binary can remain hidden in an optically normalized spectrum while producing measurable wavelength-dependent differences in the infrared.

\section{Conclusions}\label{sec:conclusions}

We have identified HD~158485 as a double-pulsator binary using pulsation timing. Six coherent $\delta$~Scuti modes separate into two groups with opposite light-travel-time variations described by a common Keplerian orbit. The direct forward model gives $\porb=382.80^{+0.80}_{-0.76}$~d, $e=0.242^{+0.045}_{-0.046}$, and a timing-based mass ratio of $q=0.818^{+0.046}_{-0.045}$. The identification is supported by a blind prediction test: an orbit fitted using only the first two long TESS campaigns predicts the third campaign without updating any orbital parameter, with $r=0.983$.

The pulsation-timing result provides the first orbit and identifies both components as pulsators, building on earlier evidence for unresolved multiplicity. Gaia DR3 had already classified the source as a likely binary, and the Hipparcos--Gaia EDR3 proper-motion anomaly provided evidence for unresolved motion. The predicted sub-20-mas separation is consistent with the non-detection in earlier speckle observations. The modest RUWE, the DR3 astrometric-binary input selection, photocentre suppression, and covariance of a near-annual orbit with parallax may help explain the absence of a published Gaia DR3 NSS orbit. The comparable optical flux contributions of the two components allow the composite spectrum to remain largely degenerate with a single-star atmosphere model.

A conditional two-atmosphere calculation gives median binary-minus-single
differences of about 1.6\% over 1--5~$\mu$m and 2.5\% over 5--28~$\mu$m. The latter is slightly above the nominal 2\% prediction accuracy targeted for an individual JWST standard \citep{Gordon2022}. This remains a conditional binary-versus-single model comparison rather than evidence for a calibration error, but it shows that unresolved multiplicity can be relevant at the accuracy level targeted for individual calibration standards.

HD~158485 therefore provides a rare example of a calibration-relevant star for which Gaia classification and the Hipparcos--Gaia proper-motion anomaly had already indicated unresolved multiplicity, but no orbit was known; TESS reveals two independent pulsation clocks and measures that orbit. Pulsation timing therefore provides a complementary way to identify and characterize unresolved binaries among calibration standards with
stable, coherent $\delta$~Scuti pulsations.

\begin{acknowledgments}
We are grateful to the anonymous reviewer for the professional and thorough suggestions. Chenglong Lv acknowledges support from the National Natural Science Foundation of China (NSFC; No. 12503042) and the Natural Science Foundation of Xinjiang Uygur Autonomous Region (No. 2025D01B176). Ali Esamdin is supported by the Tian-shan Talent Training Program (No. 2023TSYCLJ0053). This work is also supported by the Xinjiang Uygur Autonomous Region Introduced Project "Tianchi Talent".

This paper includes data collected with the \textit{TESS} mission, obtained from the Mikulski Archive for Space Telescopes (MAST) at the Space Telescope Science Institute (STScI). Funding for the \textit{TESS} mission is provided by the NASA Explorer Program. STScI is operated by the Association of Universities for Research in Astronomy, Inc., under NASA contract NAS~5-26555. The specific TESS SPOC 2-min observations analyzed in this work can be accessed from MAST via \dataset[doi:10.17909/ct8x-6583]{https://doi.org/10.17909/ct8x-6583}. This work also presents results from the European Space Agency (ESA) space mission Gaia. Gaia data are being processed by the Gaia Data Processing and Analysis Consortium (DPAC). Funding for the DPAC is provided by national institutions, in particular the institutions participating in the Gaia MultiLateral Agreement (MLA). The Gaia mission website is \url{https://www.cosmos.esa.int/gaia}. The Gaia archive website is \url{https://archives.esac.esa.int/gaia}. This work made use of the HST CALSPEC reference spectra distributed by MAST, \dataset[CALSPEC]{\doi{10.17909/t9-khb7-4049}}.

\end{acknowledgments}

\software{
Astropy \citep{Astropy2013,Astropy2018,Astropy2022},
matplotlib \citep{Hunter2007},
NumPy \citep{Harris2020},
SciPy \citep{Virtanen2020},
pandas \citep{McKinney2010},
Lightkurve \citep{Lightkurve2018},
Maelstrom \citep{Hey2020},
exoplanet \citep{ForemanMackey2021},
PyMC3 \citep{Salvatier2016},
ArviZ \citep{Kumar2019},
corner \citep{ForemanMackey2016}
}

\appendix
\twocolumngrid

\section{Validation and Posterior Diagnostics}\label{app:validation}

\begin{figure}
 \centering
 \includegraphics[width=0.45\textwidth]{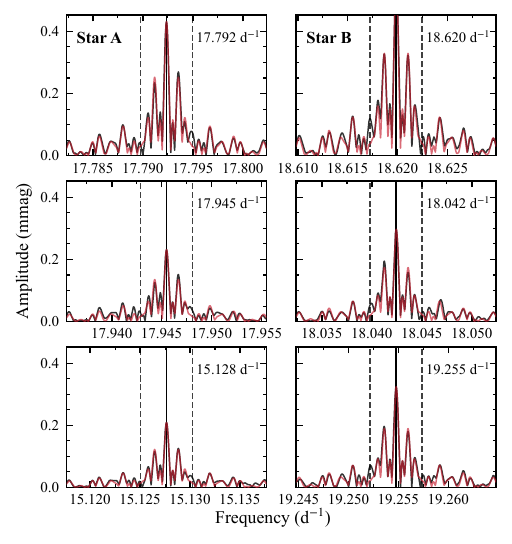}
 \caption{Local amplitude spectra around the six timing frequencies. The black lines show the observed amplitude spectra, and the red lines show the spectral window calculated from the actual TESS timestamps and scaled to the central peak in each panel. Solid black vertical lines mark the adopted frequencies, while dashed black vertical lines indicate \(\nu\pm P_{\rm orb}^{-1}\) as a frequency-scale reference. The left and right columns correspond to the two timing groups.}
 \label{fig:sidebands}
\end{figure}

The local spectra of the six timing modes are shown in Figure~\ref{fig:sidebands}. The close peaks around all six modes are reproduced by the spectral window calculated from the actual TESS timestamps; the median correlation between the
observed local profiles and the sampling-window profiles is 0.958. The dashed
lines mark $\nu\pm P_{\rm orb}^{-1}$ only as a frequency-scale reference. We therefore attribute the close structure to TESS-window aliases rather than rotational splitting.

The 20--40-d segment lengths give nearly identical periods (Table~\ref{tab:segments}). The 50-d segments partially smooth the orbital variation, producing slightly smaller light-time amplitudes and a modest shift in the recovered period. Repeating the 40-d analysis with SAP rather than PDCSAP gives a period of 386.91~d
instead of 387.01~d in the period scan.

The anti-phase relation between the two timing groups was examined independently within each TESS observing campaign. The two groups consistently show opposite delay variations across the available campaigns, as expected from two pulsating components sharing a common orbit. This campaign-by-campaign consistency supports an orbital origin of the timing signal rather than a common instrumental variation.

Two nearby control stars, TIC~232625701 and TIC~359678383, were analyzed using the same segmented timing procedure. Neither shows a comparable long-period timing variation, disfavoring a sector-dependent instrumental origin of the measured delays.

\begin{figure}
 \centering
 \includegraphics[width=0.45\textwidth]{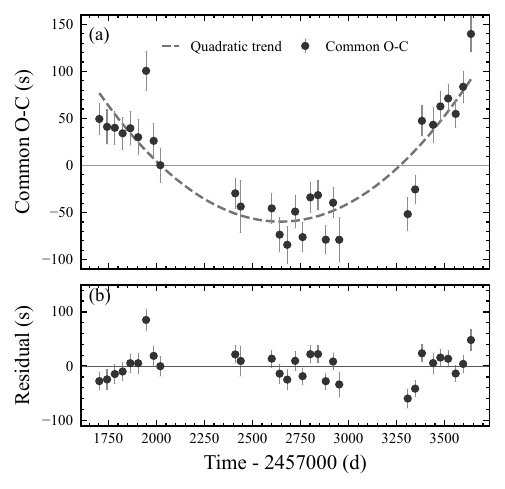}
 \caption{Common residual after subtracting the forward-model orbit. (a) The black points show the inverse-variance weighted common residuals with formal uncertainties; the gray dashed curve shows a phenomenological quadratic description of the shared residual trend. (b) Residuals after subtracting the quadratic trend.}
 \label{fig:common}
\end{figure}

The residuals after subtracting the adopted forward-model orbit contain a trend shared by the two timing groups. Figure~\ref{fig:common} shows a simple quadratic description of this residual structure. This common-mode fit was performed after subtracting the adopted forward-model orbit and was not included in the orbital likelihood. The available baseline is insufficient to establish the origin of the trend, and we do not interpret it as evidence for a third body.

\begin{figure*}
 \centering
 \includegraphics[width=0.96\textwidth]{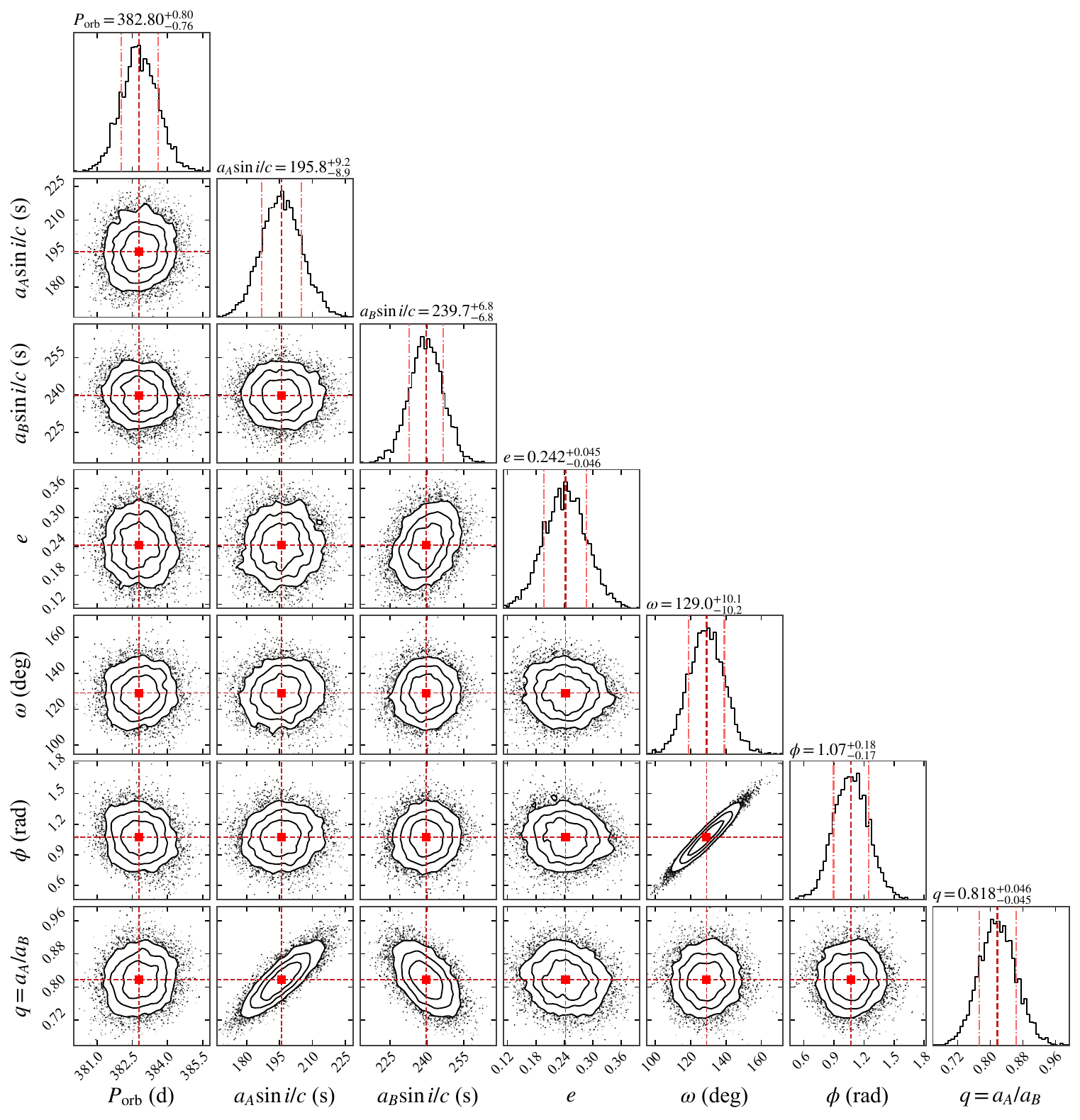}
 \caption{Posterior distributions of the forward-model orbital parameters. Red lines mark posterior medians; dashed red lines mark the 16th and 84th percentiles.}
 \label{fig:corner}
\end{figure*}

The adopted forward-model posterior is shown in Figure~\ref{fig:corner}. Among the orbital quantities, the maximum $\widehat{R}$ is 1.01 and the minimum bulk effective sample size is 483.

\begin{figure*}
 \centering
 \includegraphics[width=0.94\textwidth]{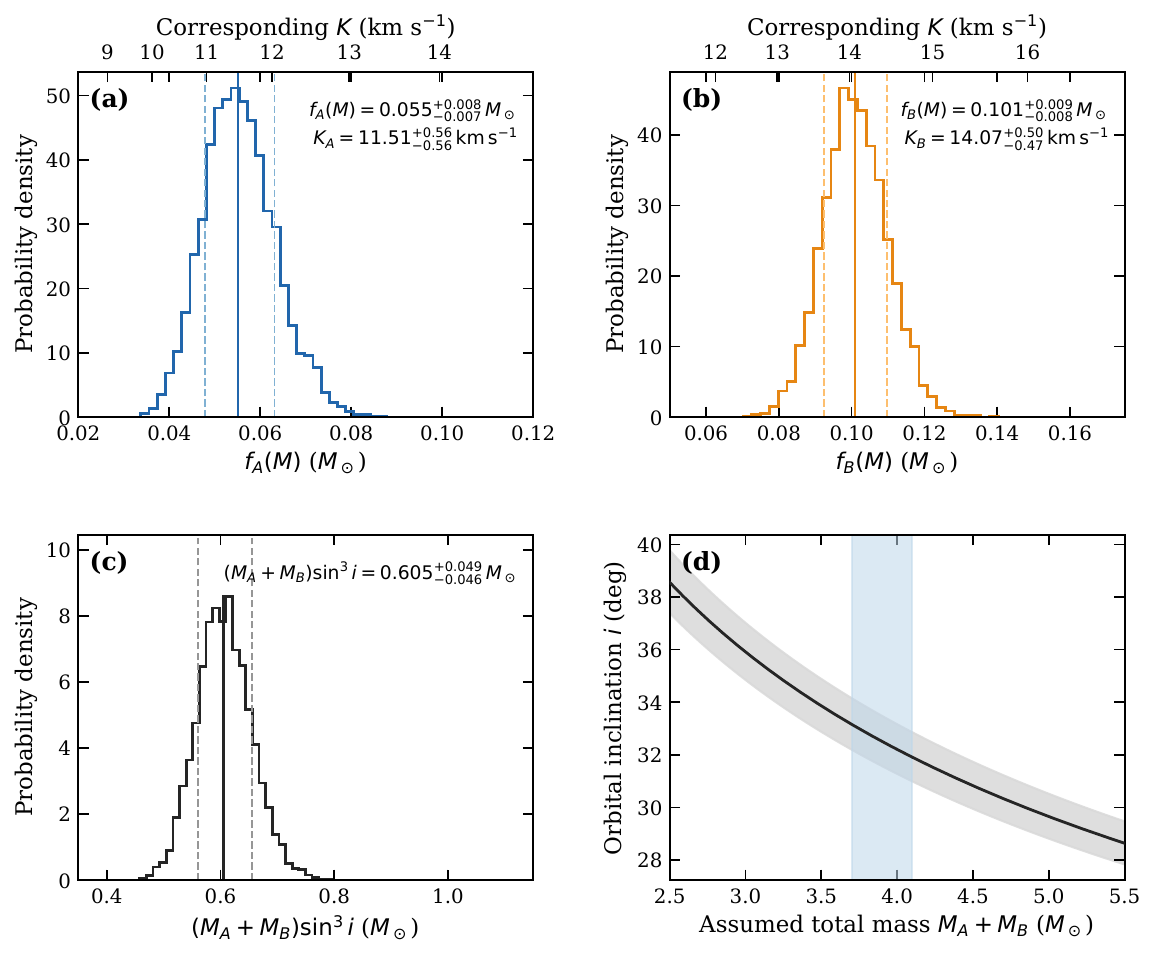}
 \caption{Physical constraints derived from the timing posterior. (a) and (b) Component mass functions with the corresponding velocity semiamplitudes on the upper axes. (c) Projected total mass, $(M_A+M_B)\sin^3 i$. Solid lines mark posterior medians and dashed lines mark the 16th and 84th percentiles. (d) Orbital inclination as a function of the assumed total system mass; the gray shaded region shows the propagated timing-posterior uncertainty. The vertical pale-blue band marks the conditional total-mass range of $3.7$--$4.1\,M_\odot$ inferred from the external component-mass estimates.}
 \label{fig:physical}
\end{figure*}

The physical constraints derived from the timing posterior are collected in Figure~\ref{fig:physical}. The mass functions, velocity semiamplitudes, and projected total mass do not require an assumed inclination. The inclination shown there is conditional on the external total-mass scale and is therefore not a direct timing measurement.

\begin{deluxetable}{ccccc}
\tablecaption{Dependence of Timing Results on Segment Length\label{tab:segments}}
\tablehead{
\colhead{Segment (d)} &
\colhead{$N_{\rm seg}$} &
\colhead{$P_{\rm orb}$ (d)} &
\colhead{$a_A\sin i/c$ (s)} &
\colhead{$a_B\sin i/c$ (s)}}
\startdata
20 & 58 & 386.98 & 190.1 & 228.8 \\
30 & 39 & 386.97 & 185.6 & 225.4 \\
40 & 30 & 387.01 & 191.8 & 222.4 \\
50 & 25 & 384.43 & 172.3 & 209.9 
\enddata
\tablecomments{These values come from the period scan of the observed segmented delays, not from separate posterior fits.}
\end{deluxetable}

\clearpage

\bibliography{references}
\bibliographystyle{aasjournalv7}

\end{document}